\documentclass[pra,twocolumn,showpacs,superscriptaddress]{revtex4}
\usepackage{graphicx}
\usepackage{amsfonts,amsmath,amssymb,float}
\usepackage{bm,dsfont}
\usepackage{color}
\usepackage{bbm}
\usepackage{longtable}
\usepackage[dvips]{epsfig}
\usepackage{amsmath,amssymb,lscape,float}
\usepackage{hyperref}
\usepackage{listings}

\begin{document}
\title{Feasibility of single-shot realizations of conditional 
three-qubit gates in exchange-coupled qubit arrays with local 
control}

\author{Vladimir M. Stojanovi\'c}
\affiliation{Department of Physics, Harvard University,
17 Oxford Street, Cambridge, Massachusetts 02138, USA} 
\affiliation{Department of Physics, University of Belgrade, 
Studentski Trg 12, 11158 Belgrade, Serbia}

\date{\today}

\begin{abstract} 
We investigate the feasibility of single-shot Toffoli- and Fredkin-gate realizations   
in qubit arrays with Heisenberg-type exchange interactions between adjacent qubits. 
As follows from the Lie-algebraic criteria of controllability, such an array is rendered completely 
controllable -- equivalent to allowing universal quantum computation -- by a Zeeman-like control 
field with two orthogonal components acting on a single ``actuator'' qubit. Adopting this local-control 
setting, we start our analysis with piecewise-constant control fields and determine the global maxima 
of the relevant figure of merit (target-gate fidelity) by combining the multistart-based clustering algorithm 
and quasi-Newton type local optimization. We subsequently introduce important practical considerations, such as 
finite frequency bandwidth of realistic fields and their leakage away from the actuator. We find the shortest times 
required for high-fidelity Toffoli- and Fredkin-gate realizations and provide comparisons to their respective 
two-qubit counterparts -- controlled-NOT and exponential-SWAP. In particular, the Toffoli-gate time compares 
much more favorably to that of controlled-NOT than in the standard decomposition-based approach. This study 
indicates that the use of the single-shot approach can alleviate the burden on control-generating 
hardware in future experimental realizations of multi-qubit gates.
\end{abstract}

\maketitle
\section{Introduction}
Recent years have witnessed rapid progress in the realm of quantum computing (QC), 
along with the development of methods for coherent control of a broad class of relevant 
systems~\cite{QCreviews2013}. With single-qubit measurement- and control 
fidelities being already above the fault-tolerance threshold, the attention 
of workers in the field is now shifting to implementing scalable qubit-array architectures 
required for fault-tolerant QC~\cite{recentQCimport}. Current developments 
in this direction are an important stride towards the establishment of noisy intermediate-scale 
quantum technology (systems containing from 50 to a few hundred qubits)~\cite{PreskillNISQ:18} 
in the next few years, a prerequisite for reaching the goal of large-scale 
universal QC (UQC). In particular, an array of nine spin qubits of Loss-DiVincenzo 
type~\cite{Loss+DiVincenzo:98,Kloeffel+Loss:13} -- with Heisenberg-type exchange interaction 
between adjacent qubits -- has already been deployed~\cite{Zajac+:16}. Arrays with this type 
of two-qubit coupling have also been realized in other physical platforms~\cite{Veldhorst+:15,Rassmussen+:18}.

The usefulness of Heisenberg interaction within the circuit model of QC has long been amply 
appreciated, despite the early realization that this interaction by itself -- unlike its lower-symmetry 
Ising and $XY$ counterparts -- does not allow for UQC~\cite{DiVincenzo+:00}. 
Importantly, it was demonstrated that UQC can still be realized with Heisenberg interaction 
alone if encoded qubit states are introduced, so that the role of logical qubits is played by 
triples~\cite{DiVincenzo+:00} or pairs~\cite{Levy:02} of physical qubits. 
This led to the concept of encoded universality~\cite{EncodedUniversality}. 
More recently, another example for the versatility of this type of two-qubit coupling was 
unravelled through Lie-algebraic studies of spin-$1/2$ systems with time-independent interacting 
Hamiltonians, which are subject to external time-dependent control fields coupled to certain 
internal degrees of freedom~\cite{Schirmer++:08,Wang++:16}. Namely, 
it was shown that a qubit array with ``always-on'' Heisenberg interaction is rendered 
completely controllable provided that at least two noncommuting controls -- for example, 
a Zeeman-type control Hamiltonian that corresponds to a magnetic field with nonzero 
components in two mutually orthogonal directions (e.g., $x$ and $y$) -- act on a single 
qubit in the array~\cite{Wang++:16}. In other words, an arbitrary 
quantum gate on any subset of qubits within the given array can then be enacted, 
which amounts to UQC. Such scenario, with external fields acting on a single 
qubit in an array, is the extreme version of the local-control approach~\cite{Heule+++}.

Regardless of the physical realization and its attendant type of two-qubit 
coupling, one of the central challenges on the way to a large-scale UQC is to reach sufficient 
accuracy in realizing quantum gates for fault-tolerant QC~\cite{PreskillNISQ:18}. A complex quantum circuit -- for 
instance, a multi-qubit gate -- can always be decomposed into a sequence of primitive single- and two-qubit 
gates~\cite{NielsenChuangBook}. Yet, such an approach is often impractical due to prohibitively 
long operation times. Besides, the number of gates needed for a quantum algorithm grows rapidly with the 
system size, with errors being accumulated with each successive gate. One alternative 
to the established decomposition-based approach entails the use of external control fields to enable fast 
{\em single-shot} realizations of multi-qubit gates. Control-based protocols~\cite{Green+:13,ControlReviews} 
have proven to be a viable route to optimized quantum-gate operations~\cite{Ashhab+:12} in 
systems ranging from superconducting~\cite{SCqubitControl,StojanovicToffoli:12,Zahedinejad++} 
to nuclear-spin-based qubit arrays~\cite{ControlNuclSpins}.

This paper investigates the feasibility of single-shot realizations of two conditional three-qubit 
gates in qubit arrays with Heisenberg interaction. To be more specific, it is focussed on quantum Toffoli 
and Fredkin gates facilitated by a Zeeman-type local control, this choice of gates being motivated by their 
importance in quantum information processing~\cite{ToffoliPapers:2017}. These two gates play key 
roles in reversible computing, each of them forming a universal gate set together with the (single-qubit) 
Hadamard gate~\cite{NielsenChuangBook}. The Toffoli gate is also an important ingredient in quantum error 
correction (QEC)~\cite{Reed++:12}.

The implementation of the Toffoli gate was already attempted in a variety of QC platforms using the standard 
decomposition-based approach~\cite{Monz++:09,Lanyon++:09,Reed++:12}. Yet, all those attempts resulted in sub-threshold 
fidelities, ranging from $68.5\%$ in circuit QED to $81\%$ in photonic systems~\cite{Lanyon++:09}. As regards 
the Fredkin gate, the progress on the experimental side is even less satisfactory. This gate was demonstrated 
non-deterministically not so long ago in linear-optics experiments~\cite{FredkinGate++}, followed by the realization 
with a fidelity of around $68\%$ in the context of entangling continuous-variable bosonic modes in three-dimensional 
circuit QED quite recently~\cite{EngExchange}.

While a number of proposals for realizing Toffoli and Fredkin gates were put forward in recent years, 
these gates have never been demonstrated in qubit arrays with Heisenberg-type interaction. 
The principal rationale for realizing them in a single-shot fashion stems from the fact that, 
e.g., for spin qubits it has proven challenging to simultaneously achieve fast, high-fidelity 
single- and two-qubit gates~\cite{Kloeffel+Loss:13}. This state of affairs is the main motivation 
for the present investigation. While this study is also motivated by the recent physical 
implementations of Heisenberg-coupled qubit arrays, here we aim for generality and thus opt 
for an implementation-independent investigation. 

The most widely used approaches presently used in high-dimensional optimization problems
entail gradient-based greedy algorithms for local optimization, which scale favorably with 
the problem size~\cite{NRcBook}. Here we determine the global maxima of the relevant figure 
of merit (gate fidelity) by combining a greedy quasi-Newton type local-optimization 
technique and the multistart-based clustering algorithm which facilitates searches for global 
maxima of objective functions~\cite{TornZilinskasBook}. In this manner, we find both 
the shortest times required for high-fidelity realizations of the chosen conditional 
three-qubit gates and the corresponding optimal control fields. 

The remainder of this paper is organized as follows. To set the stage, Sec.~\ref{local_control} 
recapitulates the main Lie-algebraic results of quantum operator control, introduces the concept 
of local control and, finally, explains its consequences for qubit arrays with Heisenberg-type 
interactions. In Sec.~\ref{System} we specify the system under investigation, describe its possible 
physical realizations, and introduce our control objectives. Sec.~\ref{Methods} is set aside for the 
description of our envisioned control scheme, as well as our procedures for finding optimal piecewise-constant 
control fields and their spectral filtering. The main findings of the paper are presented and discussed 
in Sec.~\ref{Results}. After an outlook on open-system effects in Sec.~\ref{outlook}, we conclude with 
a brief summary of the paper in Sec.~\ref{sumconclude}.
\section{Local control in qubit arrays with Heisenberg interaction}\label{local_control}
\subsection{Lie$-$algebraic criteria of controllability}\label{basics}
Consider a quantum system with a time-independent drift Hamiltonian $H_0$,
which is acted upon by external control fields $f_j(t)$ ($j=1,\ldots,p$) that 
couple to certain degrees of freedom of the system represented by Hermitian 
operators $H_j$. Its total Hamiltonian reads
\begin{equation}\label{General_Hamiltonian}
H(t)=H_0+\sum_{j=1}^{p}f_j(t)H_j \:.
\end{equation}
The dynamical equation for the time-evolution operator 
of the system, along with its initial condition, has the 
form characteristic of bilinear control systems (for convenience, 
hereafter we set $\hbar=1$)~\cite{D'AlessandroBook}:
\begin{equation}\label{time_evolution}
\dot{U}(t) = -i[H_0 +\sum_{j=1}^{p}f_j(t)H_j]\:U (t)\:, \quad U(0) =\mathbbm{1}_{n\times n}\:.
\end{equation}
The goal of a typical quantum-control problem is to find a time $t_f > 0$ and controls 
$f_j(t)\in \mathbbm{R}$ such that a desired unitary operation $U_{\textrm{target}}$ is 
reached at $t=t_f$, i.e., $U(t=t_f)=U_{\textrm{target}}$. In particular, the system is 
completely controllable if its dynamics governed by $H(t)$ can give rise -- through 
appropriately chosen fields $f_j(t)$ -- to an arbitrary unitary operation on its Hilbert 
space $\mathcal{H}$ ($n=\textrm{dim}\:\mathcal{H}$), i.e., if the reachable set of the 
system (the set of unitary operations achievable by varying the controls) coincides with 
the Lie group $U(n)$ or $SU(n)$~\cite{D'AlessandroBook}.

The controllability criteria for quantum systems are formulated using Lie-algebraic 
concepts~\cite{ControlConceptual}, with the dynamical Lie algebra (DLA)
of the system playing the central role~\cite{D'AlessandroBook}. For a system 
described by the Hamiltonian in Eq.~\eqref{General_Hamiltonian}, the DLA
$\mathcal{L}$ is generated by the operators $\{-iH_k|k=0,\ldots,p\}$, i.e., the 
skew-Hermitian counterparts of $H_k$. Importantly, a necessary and sufficient 
condition for complete controllability is that $\mathcal{L}$ is isomorphic to
$u(n)$ or $su(n)$~\cite{D'AlessandroBook}, the Lie algebras of skew-Hermitian, or 
traceless skew-Hermitian, $n\times n$ matrices~\cite{PfeiferBook}. This last result 
(the Lie-algebraic rank condition) is an existence theorem guaranteeing that any 
unitary operation on the Hilbert space of the system is reachable by an appropriate 
choice of control fields. An altogether separate question pertains to finding their 
actual time dependence that allows one to realize a desired unitary operation, taking 
into account various practical constraints such as the one on the total duration of
the control. 
\subsection{Local control and its application to qubit arrays} \label{Loc_Control}
The central control-related question in the context of interacting quantum systems
is whether a given system can be partially -- in the sense of allowing the realization 
of specific unitaries -- or, perhaps, fully controlled by solely acting on its subsystem. 
This is the principal idea behind the {\em local-control} approach. Namely, even if 
controls act only on a small subsystem, their effect may be rendered global by the 
presence of interactions. Needless to say, the choice of the relevant 
subsystem depends on the type of interaction in the physical system under
consideration.

To formalize these last considerations, assume that a composite system 
$S=C\cup \bar{C}$ is described by a Hamiltonian $H_S+\sum_j\:f^{C}_j(t) H^{C}_j$, 
where $H_S$ is the coupling (drift) part (acting on the whole $S$), and $H^{C}_j$ 
are local operators (acting on $C$ only); $f^{C}_j(t)$ are time-dependent control 
fields. Assuming, for simplicity, that $H^{j}_C$'s are generators of the local Lie 
algebra $\mathcal{L}(C)$ of skew-Hermitian operators acting on $C$, the system $S$ 
is completely controllable if and only if $iH_S$, $iH^{C}_j$ are the generators of 
its corresponding Lie algebra $\mathcal{L}(S)$, i.e., 
\begin{equation}
\langle iH_S,\mathcal{L}(C)\rangle=\mathcal{L}(S) \:,
\end{equation}
where $\langle A,B\rangle$ is the algebraic closure of the operator sets $A$ and $B$~\cite{PfeiferBook}. 
Thus, an arbitrary unitary on $S$ can be enacted via a control on its subsystem $C$ if and 
only if all elements of $\mathcal{L}(S)$ can be obtained as linear combinations of $iH_S$, 
$iH^{C}_j$, and their repeated commutators.

\begin{figure}[b!]
\includegraphics[clip,width=0.3\textwidth]{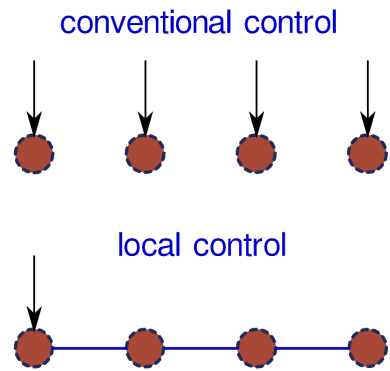}
\caption{\label{fig:QA}(Color online) Pictorial illustration of two different approaches to quantum 
control in qubit arrays: the conventional approach (top), and that of local control -- of interest 
for nearest-neighbor Heisenberg interactions -- where a single qubit in the array is acted 
upon by control fields (bottom). The circles represent qubits, while the arrows indicate 
the action of controls. The line connecting adjacent qubits in the lower part of the figure 
underscores the more prominent role that interactions play in the local-control case.}
\end{figure}

In addition to its conceptual importance, the local-control approach lends itself to applications 
in qubit arrays~\cite{LocalPlastina}, systems that provide a natural setting for UQC. 
In accordance with the above Lie-algebraic criteria (cf. Sec.~\ref{basics}), complete 
controllability of a $N$-qubit array on its $2^N$-dimensional Hilbert space requires that 
its attendant DLA be isomorphic with $u(2^N)$ or $su(2^N)$. Conventional control in an 
array entails control fields acting on each qubit to enable single-qubit operations 
(for illustration, see the upper portion of Fig.~\ref{fig:QA}). 
Combined with a drift Hamiltonian, i.e., two-qubit interactions, this 
allows in principle the realization of arbitrary (multi-qubit) gates. By contrast, in 
the local-control approach such fields act only on a small subset of {\em actuator} 
qubits, which in the extreme case can be reduced to a single qubit (the lower part of 
Fig.~\ref{fig:QA}). The choice of actuators should ideally be one that guarantees 
complete controllability, as the latter is equivalent to UQC~\cite{D'AlessandroBook}. 

Apart from its simpler implementation, local control is advantageous because reducing the number 
of controls alleviates the debilitating effects of noise and decoherence (see Sec.~\ref{outlook}). 
While not being the most critical issue in relatively small qubit arrays that are currently being 
deployed, this will become important already in the near future with the anticipated realization 
of systems with a few hundred qubits~\cite{PreskillNISQ:18} en route to large-scale UQC.
\subsection{Controllability of spin-$1/2$ chains (qubit arrays) with Heisenberg-type 
interactions}\label{Heisenberg_control}
The problem of identifying the minimal resources for controllability in 
interacting spin-$1/2$ systems, i.e., the smallest subsystem that -- when 
acted upon by external control fields -- renders the whole system completely controllable, 
was studied extensively~\cite{Schirmer++:08,Wang++:16}.
Those studies mostly relied upon standard (two-body) interacting spin-$1/2$ 
models (Ising, $XY$, Heisenberg) as their drift Hamiltonians [$H_0$ in Eq.~\eqref{General_Hamiltonian}], 
with the role of controls [$H_j$ in Eq.~\eqref{General_Hamiltonian}] 
played by local operators representing individual spins. Because these
ingredients coincide with those typically found in qubit arrays, the 
above control studies have far-reaching implications for the latter.

The most important controllability-related result for our present purposes, derived using 
a graph-infection criterion, is that the existence of two mutually noncommuting local 
controls acting on one end spin of a nearest-neighbor $XXZ$-Heisenberg spin-$1/2$ 
chain ensures complete controllability of the chain~\cite{Wang++:16}. The underlying 
$XXZ$ drift Hamiltonian reads
\begin{equation}\label{H_XXZ}
H_{XXZ}=J\sum_{i=1}^{N-1}\:\left(S^{x}_{i}S^{x}_{i+1}+S^{y}_{i}S^{y}_{i+1}
+\Delta S^{z}_{i}S^{z}_{i+1}\right) \:,
\end{equation}
\noindent where $J$ is the exchange constant and $\Delta$ the anisotropy parameter,
while the control Hamiltonian 
\begin{equation}\label{controlham}
H_c(t)=h_x(t)S^{x}_{1}+h_y(t)S^{y}_{1}
\end{equation} 
corresponds to a Zeeman-type control field $\mathbf{h}(t)\equiv[\:h_x(t),h_y(t),0\:]$ 
with nonzero $x$- and $y$ components. 

The problem of controllability in spin-$1/2$ chains (qubit arrays) with 
Heisenberg-type interactions has recently been revisited based on a method that makes use of the 
Hilbert-space decomposition into a tensor product of minimal invariant subspaces~\cite{Wang++:16}. 
In this manner, it was demonstrated that the last result holds even for the fully 
anisotropic $XYZ$ coupling case, i.e., for the drift Hamiltonian
\begin{equation}\label{H_XYZ}
H_{XYZ}=\sum_{i=1}^{N-1}\:\left(J_x S^{x}_{i}S^{x}_{i+1}+
J_y S^{y}_{i}S^{y}_{i+1}+J_z S^{z}_{i}S^{z}_{i+1}\right) \:.
\end{equation}
Moreover, it was shown that the two noncommuting controls need not be applied 
to one of the end qubits. Finally, the result holds even if the two controls are 
applied to different -- rather than the same -- qubits. Needless to say, the described 
general controllability result applies in the special case of the isotropic Hamiltonian
\begin{equation}\label{H_isotropic}
H_{XXX}=J\sum_{i=1}^{N-1}\:\left(S^{x}_{i}S^{x}_{i+1}+
S^{y}_{i}S^{y}_{i+1}+S^{z}_{i}S^{z}_{i+1}\right) \:,
\end{equation}
which is of most relevance for applications in qubit arrays. The last
Hamiltonian is the $\Delta=1$ case of Eq.~\eqref{H_XXZ}, and $J_x=J_y=J_z=J$
case of Eq.~\eqref{H_XYZ}. It will be our working drift Hamiltonian in the 
following.

For any of the Hamiltonians in Eqs.~\eqref{H_XXZ}-\eqref{H_isotropic} 
and an arbitrary fixed qubit-array size $N$, complete controllability can be 
demonstrated by showing that the DLA $\mathcal{L}_{xy}$ of the system, 
generated by the set of skew-Hermitian traceless operators 
$\{iH_d,iS^{x}_{1},iS^{y}_{1}\}$ (where $H_d=H_{XXZ},H_{XYZ}$, or $H_{XXX}$), 
has the dimension $n^2-1$, with $n\equiv 2^{N}$ being the dimension of 
the Hilbert space of the system. This implies that $\mathcal{L}_{xy}$ 
is isomorphic with $su(n)$~\cite{PfeiferBook}.

As already mentioned in Sec.~\ref{Loc_Control}, complete controllability amounts 
to UQC, i.e., it guarantees that an arbitrary gate can be enacted through
an appropriate (gate-specific) choice of control fields. 
It should be stressed that in Heisenberg-coupled qubit arrays a smaller 
degree of control than that required for UQC can be sufficient 
for nontrivial computational tasks. Namely, a single local control (e.g., an 
$x$-only Zeeman-type control on one qubit) is sufficient for controllability 
of a qubit array on all of its invariant subspaces~\cite{Wang++:16}. The largest 
among them has the dimension that is exponential in the number of qubits, thus 
being a useful quantum-computing resource. This reduced degree of control 
also allows for the realization of nontrivial gates, such as the $\mathrm{SWAP}^{1/2}$ 
(the square root of a SWAP gate) -- a natural entangling two-qubit 
gate for exchange-coupled qubits~\cite{Heule+++}.  

For completeness, it is worth mentioning that -- by contrast to those of Heisenberg-type -- other 
drift Hamiltonians of interest in realistic qubit arrays do not lead to complete controllability 
under the same (local-control) circumstances. For a $XX$-type Hamiltonian [$\Delta=0$ case in 
Eq.~\eqref{H_XXZ}], this is intimately related to the fact that the $XX$ interaction is not 
algebraically propagating~\cite{Wang++:16}. In the case of Ising coupling controls on 
each qubit are even required for complete controllability, which amounts to the conventional-control 
scenario (recall Sec.~\ref{Loc_Control}).
\section{System and target gates}\label{System}
\subsection{Total Hamiltonian and basic assumptions}\label{System_a}
In what follows, we consider a qubit array with nearest-neighbor Heisenberg 
coupling, subject to a local control of the first qubit in the array. We express 
all frequencies and control-field amplitudes in units of the coupling strength $J$ 
(recall that $\hbar=1$). Consequently, all the relevant times are expressed in units 
of $J^{-1}$. 

We take as our point of departure the Hamiltonian $H(t)=H_d+ H_c(t)$,
with the drift part given by the isotropic-Heisenberg Hamiltonian of 
Eq.~\eqref{H_isotropic} and the control part by the Zeeman-type 
Hamiltonian of Eq.~\eqref{controlham}
\begin{eqnarray}
H_d &=&\frac{J}{4}\sum_{i=1}^{N-1}\:\left(X_i
X_{i+1}+Y_{i}Y_{i+1}+Z_i Z_{i+1}\right) \:, \label{H_0Pauli} \\
H_c(t) &=& \frac{1}{2}\:[\:h_x(t)X_1+h_y(t)Y_1\:] \:, \label{controlhamPauli}
\end{eqnarray}
both, for convenience, rewritten in terms of Pauli matrices 
[\:recall that $\mathbf{S}_i=\frac{1}{2}(X_i, Y_i, Z_i)$ for qubit $i$\:].
We will also consider the effects of a static global magnetic field in the $z$ direction, 
a situation captured by the drift Hamiltonian
\begin{eqnarray}\label{Hmd}
H_{d,m}&=& \frac{J}{4}\sum_{i=1}^{N-1}\:\left(X_i X_{i+1}+Y_{i}Y_{i+1}+Z_i Z_{i+1}\right)
\nonumber \\
&-&\frac{\Omega}{2}\sum_{i=1}^{N}\:Z_i  \:,
\end{eqnarray}
where $\Omega$ quantifies the magnetic-field strength.

Realistic control fields -- e.g., magnetic fields realized using 
micromagnets~\cite{PioroLadriere+:08} -- are never perfectly localized.
Thus, our original assumption about the control being confined to a single
qubit in an array is, strictly speaking, an idealization. A more realistic scenario 
entails a field that also affects neighboring qubits due to field leakage, a situation 
that requires a slight generalization of the Hamiltonian in Eq.~\eqref{controlhamPauli}. 
Assuming an exponential field decay away from the first qubit, the relevant control 
Hamiltonian adopts the form~\cite{Wang++:16} 
\begin{equation}\label{H_cLeak}
H^{\textrm{L}}_c(t)=\frac{1}{2}\sum^{N}_{j=1}e^{-\mu_{\scriptscriptstyle\textrm{L}}
(j-1)^2}[\:h_x(t)X_j+h_y(t)Y_j\:] \:,
\end{equation}
where the parameter $\mu_{\textrm{L}}$ measures the extent of control-field leakage. 

It is worth pointing out that the field-leakage effect does not invalidate the rationale 
for using the local-control approach. Namely, in Ref.~\onlinecite{Wang++:16} it was demonstrated 
that the subspace-controllability results are robust with respect to leakage, in that the 
invariant-subspace structure and controllability of the system remain unchanged. By extension 
these results imply that the conclusions about complete controllability remain valid in the presence 
of such leakage. 

It is pertinent to stress that -- as most gate-optimization treatments -- the present work corresponds to 
the closed-system scenario, i.e., to the unitary dynamics of the system within the open-loop coherent control 
framework. In other words, the unavoidable debilitating effects of decoherence due to an interaction of qubits 
with the environment (open-system scenario) are not explicitly taken into account. The most general analysis 
of the gate-optimization problem would, however, require one to incorporate an interaction of qubits with a 
multi-mode bosonic bath, as briefly discussed in Sec.~\ref{outlook}.
\subsection{Physical realizations}
Qubit arrays with nearest-neighbor isotropic Heisenberg exchange interactions can be realized using different physical 
platforms. While this type of interaction is a natural physical ingredient in the case of spin qubits~\cite{Kloeffel+Loss:13}, 
it is worthwhile to elaborate on how it can even be realized with superconducting (SC) systems, in which the most commonly 
occurring coupling between qubits is of $XX$ type~\cite{Wendin:17}. [Note that in the condensed-matter physics terminology 
the latter is referred to as $XY$ coupling.] SC systems, in fact, allow one to realize a more general class of Hamiltonians 
than the one in Eq.~\eqref{H_isotropic} -- namely, those of the $XXZ$ Heisenberg type [cf. Eq.~\eqref{H_XXZ}]. Here we describe 
two approaches to achieve that.

One approach is based on the observation that one-dimensional arrays of capacitively-coupled SC islands 
can effectively be described as $XXZ$ spin-$1/2$ chains~\cite{Fazio+Zant:01}. Generally speaking, the $XX$
part of their effective Hamiltonian features nearest-neighbor two-body interactions, while its $Z$ part 
also comprises contributions beyond nearest neighbors. Yet, through an appropriate choice of the junction 
capacitances, as well as the capacitances of SC islands to the back gate of the structure, the $Z$ part can effectively 
be reduced to nearest-neighbor interactions. The Josephson energy $E_J$ of the junctions that couple different islands 
plays the role of the exchange coupling constant $J$ and can be varied using a magnetic field provided that those 
junctions have the form of a dc-SQUID. The $XXZ$ anisotropy parameter $\Delta$ corresponds here to the ratio 
$E_{C}/E_{J}$, where $E_C$ is the charging energy. Thus, its different values can be realized by varying this  
ratio (in particular, $\Delta=1$ is the isotropic-Heisenberg-interaction case). One SC island in the array should 
be a qubit playing the role of an actuator, while the control fields $h_x$ and $h_y$ are determined by the Josephson 
energy of this qubit (an additional constant $h_z$ field would correspond to the gate voltage). 

Another approach for realizing $XXZ$ spin-$1/2$ chains with SC qubits was quite recently laid out in Ref.~\cite{Rassmussen+:18}.
This scheme makes use of a complex SC circuit based on an array of qubits mutually connected through coupler 
circuits. The latter either contain a Josephson junction (or a dc-SQUID) in parallel with an inductor, or -- for every 
other such coupler -- these two elements connected in parallel with an additional capacitor. While several other types 
of SC qubits (transmon, X-mon, fluxonium, etc.) could in principle be utilized in this envisioned system, 
it turns out that the most realistic parameters are obtained for capacitively-shunted flux qubits~\cite{Wendin:17}.
 
In state-of-the-art solid-state QC setups in the microwave regime (based on SC- or spin qubits)~\cite{Reilly:15,vanDijk+:18}, 
precisely-shaped control pulses are obtained using arbitrary waveform generators (AWGs), currently available 
with sub-nanosecond time resolution. To be more precise, in the conventional approach for control-pulse synthesis AWGs 
only generate a baseband signal and a desired pulse is then obtained through an upconversion to microwave 
frequencies by mixing with a carrier. However, the continuously improving sampling rates of AWGs -- currently 
approaching $100$ gigasamples per second -- now allow direct digital synthesis of microwave pulses~\cite{Ryan+:17}, 
thus obviating the need for separate microwave generators. Thus, these high-bandwidth AWGs both
allow more advanced pulse shaping and reduce the number of hardware components in QC setups.

\subsection{Control objectives (target gates)}\label{control_obj}
Our objective is to realize Toffoli and Fredkin gates in an array with $N=3$ qubits with 
the first qubit playing the role of actuator. The same qubit will also be the control qubit 
in the Fredkin-gate realizations, while in the context of the Toffoli gate it will also be 
one of the control qubits. At the same time, the third qubit will play the role of the target 
qubit for both gates.

A Toffoli (controlled-controlled-NOT) gate enacts a Pauli-$X$ (flip) operation on the third 
(target) qubit if the first two (control) qubits are both set (i.e., both are in the $|1\rangle$ 
state), doing nothing otherwise~\cite{NielsenChuangBook},. It represents a generalization of 
controlled-NOT (CNOT), an entangling two-qubit gate. Arguably the most important application of 
the Toffoli gate is in the measurement-free QEC~\cite{Reed++:12}, where it effectively replaces the 
measurement- and correction steps of the standard QEC.

A Fredkin (controlled-SWAP) gate enacts a SWAP operation between the second 
and third qubits, if the first (ancilla) qubit is set, otherwise leaving their 
states unchanged~\cite{NielsenChuangBook}. In other words, it enacts an entangling operation 
between those two qubits by performing a superposition of the identity and SWAP 
gates. While this operation is conditioned on the state of the ancilla qubit, thus 
giving rise to tripartite entanglement, its closest two-qubit counterpart is 
exponential-SWAP (eSWAP) -- an unconditional entangling operation given by 
$\exp(i\theta_c\textrm{SWAP})\equiv\cos\theta_c\mathbbm{1}_{4\times 4}+i\sin\theta_c\:\textrm{SWAP}$. 

Toffoli and Fredkin gates are self-inverse operations ($U_{\textrm{gate}}=U^{-1}_{\textrm{gate}}$), i.e., 
two consecutive applications of these gates amount to the identity operation ($U^2_{\textrm{gate}}=\mathbbm{1}$). 
This fact has profound consequences for the shape of the optimal control-pulse sequences (see Sec.~\ref{PiecewiseFiltered}).
\section{Methodology}\label{Methods}
\subsection{Control scheme and its justification} \label{control_scheme}
Our goal is to find the time dependence of control fields $h_{x}(t)$ and $h_{y}(t)$ 
for high-fidelity realizations of the desired three-qubit gates in a system with $N=3$ 
qubits. For the sake of simplicity, we will attempt to synthesize the corresponding 
optimal-field waveforms starting from piecewise-constant (hereafter abbreviated as PWC) 
control fields applied in alternation in the $x$ and $y$ directions with the respective 
amplitudes $h_{x,n}$ and $h_{y,n}$ ($n=1,\ldots,N_{f}/2$). In the following, we describe 
our envisioned control scheme, which represents one special realization of the control 
Hamiltonian in Eq.~\eqref{controlhamPauli}. 

At $t=0$ a pulse is applied in the $x$ direction with the constant amplitude 
$h_{x,1}$ during the time interval $0\leq t< T$. The corresponding Hamiltonian 
of the system during this interval is given by $H_{x,1}\equiv H_d+\:(h_{x,1}/2)X_{1}$.
Then a $y$ pulse with the amplitude $h_{y,1}$ is applied during the interval 
$T\leq t< 2T$, whereby the system dynamics are governed by the Hamiltonian 
$H_{y,1}\equiv H_d+(h_{y,1}/2)Y_{1}$. This sequence of alternating $x$ and $y$ pulses 
is continued until $N_f$ pulses have been carried out by the time $t_f\equiv N_f T$. 
The resulting time-evolution operator of the system is obtained by concatenating
operators $U_{y,n}\equiv\exp(-iH_{x,n}T)$ and $U_{x,n}\equiv\exp(-iH_{y,n}T)$ for
$n=1,\ldots,N_{f}/2$: 
\begin{equation}\label{deftimeevolop}
U(t=t_f)=U_{y,N_f/2}\:U_{x,N_f/2}\ldots U_{y,1}U_{x,1}\:.
\end{equation}

Our chosen local-control scheme is based on a successive switching between $x$- and $y$-control Hamiltonians 
($H_{x,n}$ and $H_{y,n}$, respectively). It represents a slight generalization of a well-known switching 
scheme that was proven by Lloyd to be sufficient for UQC~\cite{Lloyd:95}. Namely, if $A$ and $B$ are 
Hermitian matrices of dimension $d\ge 2$ and $\mathcal{L}$ the Lie algebra they generate through 
commutation, then for any $L\in\mathcal{L}$ the unitary matrix $U=e^{iL}$ can be expressed in the 
form $U=e^{-iBt_{2k}}e^{-iAt_{2k-1}}\ldots e^{-iBt_2}e^{-iAt_1}$ with finite $k$. This last result, 
which was put on a rigorous mathematical footing in Ref.~\onlinecite{Weaver:00}, can be viewed as a 
consequence of an even more general result pertaining to uniform finite generation of connected compact Lie groups 
[such as $U(n)$ or $SU(n)$]~\cite{D'AlessandroBook}. That result asserts that for a connected Lie group $e^{\mathcal{L}}$ 
corresponding to a Lie algebra $\mathcal{L}$, every element $U\in e^{\mathcal{L}}$ can be expressed 
through a finite number of factors of the type $e^{-iA_{r}t_r}$, where $A_{r}$ is one of the generators 
of $\mathcal{L}$ and $t_r>0$. 

The crucial implication of the above mathematical results for the system at hand is that an arbitrary unitary 
operation acting on its Hilbert space -- including our target conditional three-qubit gates -- can 
be obtained with a finite sequence of operators $\exp(-iH_{x,n}t_{x,n})$ and $\exp(-iH_{y,n}t_{y,n})$. 
For the sake of simplicity, our elected control schemes assumes that all time slices have equal length, 
i.e., that $t_{x,n}=t_{y,n}=T$ for each $n$.
\subsection{Numerical optimization of target-gate fidelities} \label{optim_method}
The problem at hand represents a unitary gate synthesis up to a global phase in a closed
quantum system. Therefore, we will make use of the standard figure of merit for this subclass 
of problems in quantum operator control -- the (normalized) phase invariant distance to the 
target unitary (here a three-qubit quantum gate) $U_{\mathrm{gate}}$ at the final time 
$t=t_f$ -- i.e., the trace fidelity
\begin{equation}\label{deffidelity}
F(t=t_f)=2^{-N}\:\big|\mathrm{Tr}\big[U^{\dag}(t=t_f)
U_{\mathrm{gate}}\big]\big|\:.
\end{equation}  
Needless to say, in accordance with the comments made at the end of Sec.~\ref{System_a}, the
last expression and all the results to be presented below correspond to the {\em intrinsic}
fidelity (fidelity in the absence of decoherence).

For each target gate, we maximize its fidelity -- equivalent to minimizing the gate error 
$1-F(t=t_f)$ -- over the control amplitudes $h_{x,n}$, $h_{y,n}$  ($n=1,\ldots,N_{f}/2$) 
for varying total number ($N_f$) and duration ($T$) of pulses (time slices). Finding the 
global maximum of $F$ constitutes a nontrivial numerical-optimization problem. We perform 
this complex task using the {\em multistart-based clustering} algorithm which entails the 
following steps~\cite{TornZilinskasBook}. One starts from a large sample of random points 
in the space of candidate solutions (set of control amplitudes). One then selects a smaller 
number of them that yield the largest fidelities and performs local searches for maxima around 
each of these points: the one with the highest value of the fidelity is then adopted as the 
sought-after global maximum. The validity of this approach is corroborated by the stability 
of the final result for the global maximum upon varying the initial number of random points.

Local searches for the maxima of the target-gate fidelity [Eq.~\eqref{deffidelity}] are performed using a 
second-order concurrent-update method of the quasi-Newton type~\cite{NRcBook}. The latter requires 
one to start from an initial guess for the values of the relevant variables (here control-field 
amplitudes) and an initial Hessian approximation (here taken to be the identity). The control 
amplitudes are then iteratively updated according to the Newton update rule, while the approximate 
Hessian is constructed from the past gradient history according to the Broyden-Fletcher-Goldfarb-Shanno 
(BFGS) formula~\cite{NRcBook}. After each iteration the objective function [here $F(t=t_f)$] increases, 
with termination when the desired accuracy is reached. This procedure guarantees convergence 
to a local maximum of the objective function.
\subsection{Spectral filtering of optimal PWC control fields} \label{smoothing}
While being a conventional starting point in gate optimizations, PWC control fields -- which have 
infinite spectral bandwidths -- represent a mathematical idealization. In realistic implementations 
of quantum control, achievable fields have nonzero Fourier components only in a finite frequency range. 
Thus, in order to make contact with experiments it is necessary to perform spectral filtering of optimal 
PWC control fields. In particular, low-pass filters are inherent to all present-day AWGs.

Constraints on the frequency spectra of the control fields $h_j(t)$
($j=x,y$) are imposed through filter functions. The filtered fields $h^{f}_j(t)$ 
are obtained by first operating with a filter function $f(\omega)$ on the Fourier 
transforms $\mathcal{F}[h_j](\omega)$ of the optimal fields, and then switching 
back to the time domain via inverse transform
\begin{equation}\label{filtering_procedure}
h^{f}_j(t)=\mathcal{F}^{-1}\big[f(\omega)
\mathcal{F}[h_j](\omega)\big] \quad (\:j=x,y\:)\:.
\end{equation}
In particular, we will make use of an {\em ideal low-pass} filter 
which removes the Fourier components at frequencies outside the interval 
$[-\omega_0,\omega_0]$, i.e., $f(\omega)=\theta(\omega+\omega_0)-\theta(\omega-\omega_0)$.
Using the general expression in Eq.~\eqref{filtering_procedure}, in this special case 
we obtain~\cite{Heule+++}
\begin{equation}\label{tildeh}
\begin{split}
h^{f}_x(t)=\frac{1}{\pi}\sum_{n=1}^{N_t/2}
h_{x,n}\big[a_{2n-1}(t)-a_{2n-2}(t)\big]\:,\\
h^{f}_y(t)=\frac{1}{\pi}\sum_{n=1}^{N_t/2}
h_{y,n}\big[a_{2n}(t)-a_{2n-1}(t)\big]\:,
\end{split}
\end{equation}
where $a_{m}(t)\equiv\mathrm{Si}\big[\omega_0(mT-t)\big]$ and
$\mathrm{Si}(x)\equiv\int_0^{x}(\sin{t}/t)dt$. The time-evolution operator 
corresponding to the filtered fields -- from which the attendant gate fidelities 
are easily obtained using the expression in Eq.~\eqref{deffidelity} -- can be computed 
using an unconditionally stable numerical method based on a product-formula 
approximation~\cite{Heule+++}. 

Therefore, experimentally-feasible (finite-bandwidth) control fields are here obtained through 
post-processing, i.e., low-pass filtering, of their optimized PWC counterparts. In connection 
with our use of this approach, the following two remarks are in order here. 

While PWC control fields successively applied in the $x$- and $y$ directions represent 
our point of departure in the present work (cf. Sec.~\ref{control_scheme}), the filtered fields $h^{f}_x(t)$ 
and $h^{f}_y(t)$ generically both have nonzero values throughout the interval $[0,t_f]$. Thus, our inherently 
simple control scheme does not bear a significant loss of generality compared to the more general control 
protocol in which the initial PWC control fields simultaneously have nonzero components in both relevant 
spatial directions. 

For completeness, it is worthwhile to note that an alternative approach to obtaining finite-bandwidth control 
fields would entail imposing a spectral-bandwidth constraint from the outset, i.e., incorporating it {\em a priori} 
in the numerical search for optimal control fields. Such an approach has quite recently been demonstrated by 
Lucarelli~\cite{Lucarelli:18}. That approach -- computationally much more demanding than the one utilized here -- relies 
on the use of Slepian sequences, finite-length sequences that represent the space of band-limited signals and 
serve as the basis functions for PWC control fields. 
\section{Results and Discussion} \label{Results}
In what follows, we present our findings, starting from the results obtained for the Toffoli- 
and Fredkin gate fidelities in the idealized situation where PWC control fields act on a single actuator 
qubit. We then discuss a more realistic setting that entails filtered control fields or their leakage 
away from the actuator. Finally, we also discuss the effect that the presence of a static global magnetic 
field has on the gate fidelities and gate times. To illustrate the efficiency of our approach, we also provide 
comparisons of the Toffoli- and Fredkin gate times with the gate times corresponding to their respective 
two-qubit counterparts -- CNOT and eSWAP.

As our target intrinsic gate fidelity for PWC control fields we adopt $1-10^{-4}$ (i.e., $99.99\%$), 
which was once widely accepted as the threshold for fault-tolerant QC~\cite{Steane:03,KnillThreshold}. 
It is worth mentioning, however, that QC schemes with significantly lower thresholds -- with gate errors 
as high as $10^{-2}$ -- have been proposed more recently~\cite{HighThreshold}. Bearing this is mind, we 
adopt $10^{-2}$ and $10^{-3}$ as our target gate errors -- corresponding to the target intrinsic gate 
fidelities of $99\%$ and $99.9\%$, respectively -- in the more realistic setting described above.
\subsection{Optimal PWC and filtered control fields}\label{PiecewiseFiltered}
Our optimization of the gate fidelities shows two central features. Firstly, for a fixed total 
time $t_f$ the gate fidelities depend significantly on the total number $N_f$ of pulses; they 
increase upon increasing $N_f$ (or, equivalently, decreasing the duration $T$ of a single 
pulse). Secondly, each gate has its own characteristic shortest gate time required to reach 
a fidelity close to unity, below which such fidelities are unreachable regardless of $N_f$.

\begin{figure}[t!]
\includegraphics[width=0.85\linewidth]{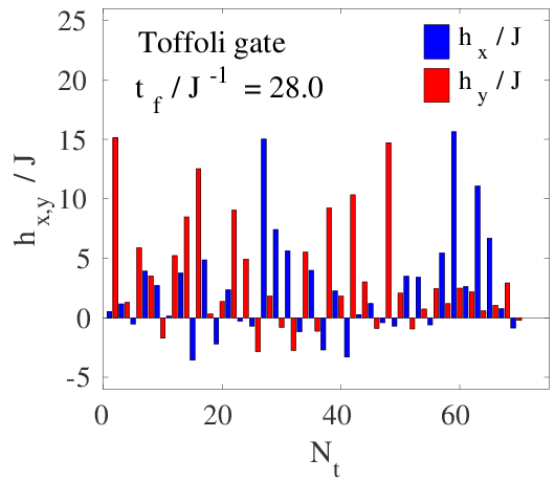}
\caption{\label{optimal_fields_Toffoli}(Color online) Optimal piecewise-constant control fields 
realizing a Toffoli gate with the fidelity $F=1-10^{-4}$, corresponding to $N_f=70$ and 
$t_{f}=28.0\:J^{-1}$.}
\end{figure}

In particular, we find that the shortest Toffoli-gate time required
to reach an intrinsic fidelity $F=1-10^{-4}$ within our approach is 
approximately $t_{f}=28\:J^{-1}$. The corresponding optimal $x$- and 
$y$ PWC control fields, corresponding to $N_f=70$ pulses, 
are depicted in Fig.~\ref{optimal_fields_Toffoli}. The shortest times required 
to realize the same gate with somewhat larger gate errors of $10^{-3}$ and 
$10^{-2}$ are approximately $25\:J^{-1}$ and $21\:J^{-1}$, respectively.
\begin{figure}[b!]
\includegraphics[width=0.85\linewidth]{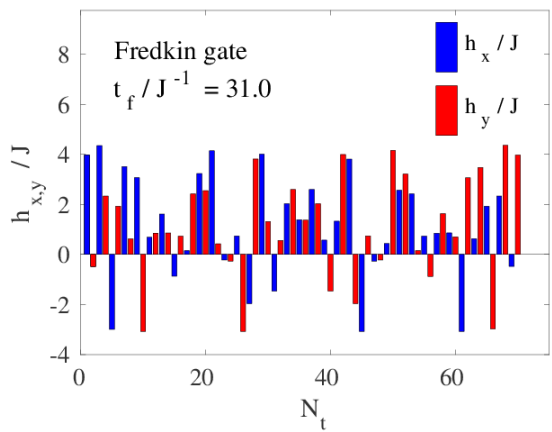}
\caption{\label{optimal_fields_Fredkin}(Color online) Optimal piecewise-constant 
control fields realizing a Fredkin gate with the fidelity $F=1-10^{-4}$, corresponding 
to $N_f=70$ and $t_{f}=31.0\:J^{-1}$.}
\end{figure}

It is instructive to compare the obtained Toffoli-gate times with those
of CNOT, its two-qubit counterpart. For instance, the shortest times 
needed to realize CNOT on the second- and third qubit in the same system with 
the respective fidelities of $1-10^{-4}$ and $1-10^{-3}$ (i.e.,the gate 
errors of $10^{-4}$ and $10^{-3}$) we find to be approximately $25.1\:J^{-1}$ and 
$17.3\:J^{-1}$. Therefore, the shortest Toffoli-gate time within our single-shot 
approach compares much more favorably to that of CNOT than is the case within the 
standard decomposition-based approach, where the optimal CNOT-gate cost of a Toffoli 
gate is 6~\cite{Shende+Markov:09}. In previous studies of single-shot gate realization, 
favorable comparisons of Toffoli- and CNOT gate times were found only in some special 
cases~\cite{Ashhab+:12}. Thus, the results obtained here can be attributed to the 
versatility of the exchange interaction and our adopted control scenario.

As regards the Fredkin gate, the shortest time required to realize it with an
intrinsic fidelity $F=1-10^{-4}$ is approximately $t_{f}=31\:J^{-1}$, while the 
respective times needed to realize this gate with the errors of $10^{-3}$ and 
$10^{-2}$ are approximately $28\:J^{-1}$ and $24\:J^{-1}$. The optimal pulse 
sequence corresponding to $F=1-10^{-4}$, with the total of $N_f=70$ pulses, is 
depicted in Fig.~\ref{optimal_fields_Fredkin} and has the interesting property 
of being palindromic in nature. 

It is worth pointing out that palindromic pulse sequences are a common occurrence for self-inverse 
gates (͑cf. Sec.~\ref{control_obj}) and result from specific properties of underlying Hamiltonians 
under the time-reversal transformation ($t\rightarrow -t$, $\mathbf{S}_i \rightarrow -\mathbf{S}_i$).
Because the Toffoli gate is a self-inverse operation too, a palindromic optimal pulse sequence 
could have, in principle, also been expected for this gate. Yet, our numerical-optimization 
procedure apparently yields another pulse sequence that corresponds to a higher fidelity.

By analogy to the previously made comparison between the Toffoli and CNOT-gate times, 
it is judicious to compare the obtained Fredkin-gate times with those corresponding 
to the closely related two-qubit gate -- eSWAP (recall Sec.~\ref{control_obj}). 
Our numerical computation shows that the shortest times required to realize the eSWAP gates 
corresponding to $\theta_c=\pi/6,\pi/4$, and $\pi/3$ with an intrinsic fidelity of $1-10^{-2}$ 
are all aproximately equal to $21\:J^{-1}$. For the eSWAP-gate times needed to reach an 
intrinsic fidelity of $1-10^{-3}$ we obtain $24\:J^{-1}$ for $\theta_c=\pi/6$ and $\pi/4$, 
while for $\theta_c=\pi/3$ we get $22\:J^{-1}$. Finally, those that correspond to $F=1-10^{-4}$ 
are approximately $28\:J^{-1}$ for $\theta_c=\pi/6$, $29\:J^{-1}$ for $\theta_c=\pi/4$, and 
$34\:J^{-1}$ for $\theta_c=\pi/3$. Thus the obtained Fredkin-gate times are just slightly 
longer than those characteristic of eSWAP, which is another sign of the effectiveness of our 
approach.

\begin{figure}[b!]
\begin{center}
\includegraphics[width=0.9\linewidth]{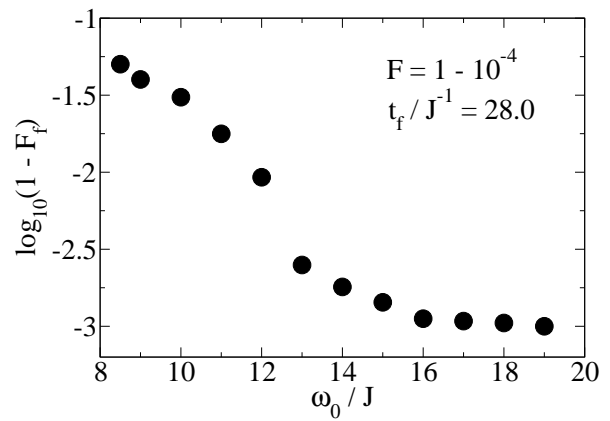}
\caption{\label{FilterToffoli}Logarithm of the gate 
error $1-F_f$ that corresponds to the low-pass filtered control fields
resulting from the optimal piecewise-constant control-pulse sequence realizing the Toffoli 
gate (cf. Fig.~\ref{optimal_fields_Toffoli}). $\omega_0$ is the cut-off frequency.}
\end{center}
\end{figure}

\begin{figure}[t!]
\begin{center}
\includegraphics[width=0.9\linewidth]{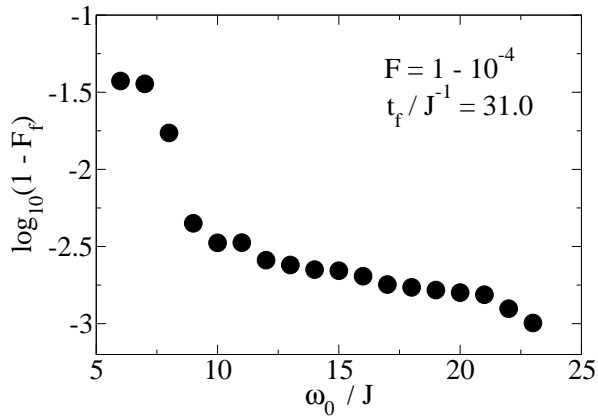}
\caption{\label{FilterFredkin}Logarithm of the gate error $1-F_f$ that corresponds 
to the low-pass filtered control fields resulting from the optimal piecewise-constant 
control-pulse sequence realizing the Fredkin gate (cf. Fig.~\ref{optimal_fields_Fredkin}). 
$\omega_0$ is the cut-off frequency.}
\end{center}
\end{figure}

The quantitative effect of spectral low-pass filtering of the obtained optimal PWC
control fields on the Toffoli- and Fredkin-gate fidelities is illustrated in Figs.~\ref{FilterToffoli} 
and \ref{FilterFredkin}, respectively, where the gate error corresponding to the 
low-pass filtered control fields is shown. What can be inferred from these plots 
is that fidelities as high as $1-10^{-3}$ can be preserved for the cut-off frequencies 
$\omega_0 \gtrsim 16\:J$ (Toffoli gate) and $\omega_0 \gtrsim 23\:J$ (Fredkin gate).

It is worthwhile to stress that for a typical range of magnitudes of exchange-coupling constants in spin- 
and SC-qubit systems ($J/2\pi\hbar \sim\:20 - 50$ MHz), the obtained cut-off frequencies are well within 
the range achievable with state-of-the-art AWGs~\cite{StojanovicToffoli:12}. Thus, low-pass filtering (recall 
Sec.~\ref{smoothing}) -- which turns infinite-bandwidth optimal PWC control fields into realistic finite-bandwidth 
ones -- does not present an obstacle to achieving high gate fidelities within our present approach.
\subsection{Effects of control-field leakage}
\begin{figure}[t!]
\begin{center}
\includegraphics[width=0.85\linewidth]{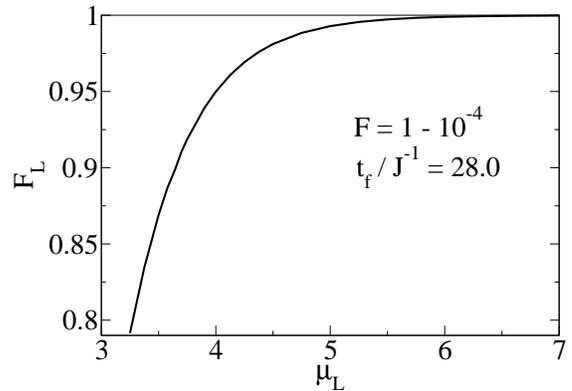}
\caption{\label{LeakToffoli}Toffoli-gate fidelity in the presence
of control-field leakage away from the actuator qubit, 
characterized by the parameter $\mu_{\textrm{L}}$. The results 
correspond to the piecewise-constant control fields optimized for
the leakage-free case (Fig.~\ref{optimal_fields_Toffoli}).}
\end{center}
\end{figure}

\begin{figure}[b!]
\begin{center}
\includegraphics[width=0.85\linewidth]{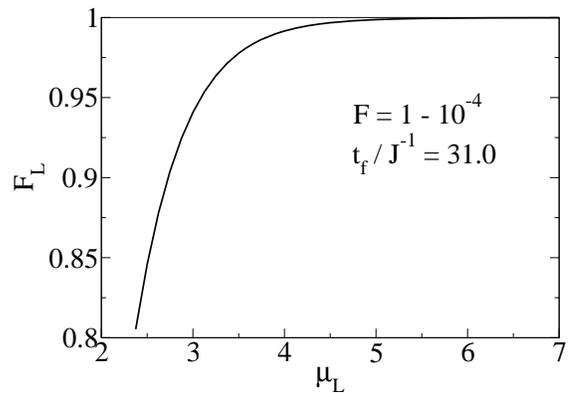}
\caption{\label{LeakFredkin}Fredkin-gate fidelity in the presence
of control-field leakage away from the actuator qubit, characterized 
by the parameter $\mu_{\textrm{L}}$. The results correspond to the 
piecewise-constant control fields optimized for the leakage-free 
case (Fig.~\ref{optimal_fields_Fredkin}).}
\end{center}
\end{figure}

The effect of control-field leakage -- as quantified by the parameter $\mu_{\textrm{L}}$ -- on 
the fidelities of the Toffoli and Fredkin gates is illustrated in Figs.~\ref{LeakToffoli} and 
\ref{LeakFredkin}, respectively. These results make it possible to draw conclusions about 
the permissible extent of leakage that allows the preservation of high gate fidelities. 
For the Toffoli gate, our calculations show that in order to retain fidelities above $1-10^{-2}$ 
($1-10^{-3}$) for the control-pulse sequences optimized for the leakage-free case one needs 
$\mu_{\textrm{L}}\gtrsim 5$ ($\mu_{\textrm{L}}\gtrsim 5.5$), implying that the magnitude of 
stray fields on the nearest neighbor of the actuator qubit does not exceed $0.7\%$ ($0.4\%$) 
of the original field. Similarly, in the case of the Fredkin gate for preserving such fidelities 
one needs $\mu_{\textrm{L}}\gtrsim 4$ ($\mu_{\textrm{L}}\gtrsim 4.5$). The corresponding magnitude 
of stray fields on the qubit adjacent to the actuator does not exceed $1.8\%$ ($1.1\%$) of the 
original field.

The obtained results for the critical extent of leakage that allows
high-fidelity realization of the chosen gates should, however, not be taken 
as a sign that the proposed single-shot approach is highly sensitive to 
the leakage effects. Namely, the curves in Figs.~\ref{LeakToffoli} and 
\ref{LeakFredkin} show the obtained results for the gate fidelities in the presence 
of leakage, but those results correspond to the control-pulse sequences optimized for 
the leakage-free case, where the relevant control Hamiltonian is the one given by 
Eq.~\eqref{controlhamPauli}. Therefore, they should merely be viewed as benchmark 
curves, to be used for extracting the actual (system- and gate-specific) value of 
the leakage parameter $\mu_{\textrm{L}}=\mu^{*}_{\textrm{L}}$. This can be done by 
comparing the relevant benchmark curve with the fidelity obtained by experimentally 
running the relevant optimal pulse sequence. 

The {\em in-situ} leakage-parameter retrieval of the kind described above should be viewed as the first 
step in any realistic application of the single-shot approach in the local-control setting. Its second 
step should entail finding another pulse sequence, this time optimized in the presence of leakage, i.e., 
assuming that the system dynamics are governed by the control Hamiltonian given by Eq.~\eqref{H_cLeak}, 
with the previously extracted value $\mu_{\textrm{L}}=\mu^{*}_{\textrm{L}}$ of the leakage 
parameter. This optimization can be carried out using the same approach as in the absence of leakage 
(cf. Sec.~\ref{optim_method}). As our explicit numerical calculations demonstrate, very high fidelities 
are achievable even for those values of $\mu_{\textrm{L}}$ whose corresponding fidelities in 
the two benchmark curves significantly deviate from unity. Interestingly, the corresponding 
gate times are similar to, and in some cases even shorter, than their counterparts in the 
leakage-free case. 

For instance, in the case of the Toffoli-gate realization with $\mu^{*}_{\textrm{L}}=3.25$, 
where the corresponding fidelity in the benchmark curve (Fig.~\ref{LeakToffoli}) is rather low, 
more precisely slightly below $0.8$, our calculation shows that the fidelity of $1-10^{-3}$ can be 
obtained within approximately $23\:J^{-1}$. This is actually a shorter gate time than that required 
for the same fidelity in the leakage-free case ($25\:J^{-1}$). Similarly, in the Fredkin-gate realization 
with $\mu^{*}_{\textrm{L}}=3.5$, where the relevant fidelity in the curve of Fig.~\ref{LeakFredkin} 
is around $0.98$, a fidelity as high as $1-10^{-4}$ can be obtained with the gate time of approximately 
$28\:J^{-1}$, significantly shorter than $31\:J^{-1}$ in the absence of leakage. These results clearly 
indicate that our proposed two-step procedure constitutes an efficient scheme for achieving high gate 
fidelities even in the presence of a substantial control-field leakage away from the actuator qubit.
While it was already stated that the presence of leakage does not invalidate the theoretical (Lie-algebraic) 
basis for the local-control approach (recall the discussion in Sec.~\ref{System_a}), our numerical 
findings strongly suggest that it also does not diminish the potential practical effectiveness of 
this approach.
\subsection{Effects of a global magnetic field}
In addition to the results obtained in the case of the isotropic-Heisenberg drift Hamiltonian 
$H_d$ of Eq.~\eqref{H_0Pauli}, it is of interest to also analyze the effect that the presence of a 
residual global magnetic field has on the gate fidelities and the corresponding gate times. This situation 
is described by the extended drift Hamiltonian $H_{d,m}$ of Eq.~\eqref{Hmd}, where the strength of a static 
Zeeman-type magnetic field in the $z$ direction is parameterized by $\Omega$. The numerical procedure utilized 
to optimize the gate fidelities over control-field amplitudes is exactly the same as in the field-free case 
(cf. Sec.~\ref{optim_method}).

\begin{table}[t!]
\begin{center}
\begin{tabular}{c|c|c|c|c}\hline
  & \multicolumn{2}{|c|}{Toffoli-gate time $[J^{-1}]$} & \multicolumn{2}{c}{Fredkin-gate time $[J^{-1}]$}\\ \hline
$\Omega/J$ & $F=1-10^{-2}$ & $F=1-10^{-3}$ & $F=1-10^{-2}$ & $F=1-10^{-3}$\\ \hline
0 & 21 & 25 & 24 & 28 \\ \hline
0.1 & 21 & 29 & 24 & 33 \\ \hline
0.2 & 21 & 27 & 25 & 34 \\ \hline
0.3 & 18 & 25 & 20 & 29 \\ \hline
0.4 & 22 & 25 & 26 & 29 \\ \hline
0.5 & 21 & 26 & 24 & 25 \\ \hline
0.6 & 19 & 27 & 19 & 30 \\ \hline
0.7 & 19 & 28 & 20 & 29 \\ \hline
0.8 & 22 & 28 & 24 & 36 \\ \hline
0.9 & 21 & 22 & 23 & 36 \\ \hline
1.0 & 19 & 30 & 19 & 31\\ \hline
1.1 & 20 & 25 & 20 & 30\\ \hline
1.5 & 18 & 27 & 23 & 29\\ \hline
\end{tabular}
\caption{Approximate Toffoli and Fredkin gate times in the presence of a 
global magnetic field quantified by $\Omega$.} \label{GlobalTable}
\end{center}
\end{table}

The approximate Toffoli- and Fredkin-gate times corresponding to the target intrinsic 
fidelities of $1-10^{-2}$ and $1-10^{-3}$, obtained for a wide range of values for $\Omega/J$, 
are summarized in Table~\ref{GlobalTable}. For both three-qubit gates under consideration 
and both stated target values of the corresponding fidelities, the obtained gate 
times show an apparent nonmonotonic behavior with increasing $\Omega/J$ and do not deviate 
significantly from their counterparts found in the absence of the external field. 
Interestingly, for the target fidelity of $1-10^{-2}$, the shortest Toffoli- and 
Fredkin gate times are quite similar and obtained for the same values of $\Omega/J$ 
($\Omega/J=0.3,0.6,1.0$). This is no longer the case for the higher target fidelity
of $1-10^{-3}$, where the shortest obtained times for the Toffoli and Fredkin gates 
correspond to different (non-zero) field strengths. 
\subsection{Comparison to other approaches for realizing conditional gates}
It is instructive to compare the present approach to realizing the Toffoli and Fredkin 
gates in Heisenberg-coupled qubit arrays to some recent related works. 

An efficient scheme has recently been proposed for realizing these conditional three-qubit 
gates in a SC circuit that comprises two qubits and one qutrit (a three-level generalization of a qubit) 
and effectively represents an $XXZ$ Heisenberg chain~\cite{AarhusCircuitGates:18}. That scheme is, in fact, 
more general and apart from those two gates can implement in principle any controlled-controlled 
unitary operation. The latter are exemplified by the double-controlled holonomic single-qubit gate, based 
on the idea of holonomic quantum computation~\cite{Zanardi+Rasetti:99} -- a general framework for building universal 
sets of robust gates using non-Abelian geometric phases. While holonomic gates were originally envisioned to be 
adiabatic, the scheme in Ref.~\onlinecite{AarhusCircuitGates:18} implements them in a non-adiabatic 
fashion~\cite{Sjoqvist+:12}. 

One obvious common denominator of the present work, based on the optimal-control theory, and the scheme proposed in 
Ref.~\onlinecite{AarhusCircuitGates:18} is their increased robustness to noise compared to the conventional control 
protocols. While here this robustness stems from the reduced number of actuator qubits 
(local control), in the latter scheme it originates from the geometric character of holonomic gates. In particular, 
nonadiabatic implementations~\cite{Feng+:13,Chen+:18} of holonomic gates generally lead to shortened gate times and thereby 
alleviate the loss of coherence (due to exposure to open-system effects) that typically hampers their adiabatic counterparts. 
The same effect that can also be achieved using an approach that became known as the shortcut to adiabaticity~\cite{Chen+:11,Ibanez+:12,Zhang+:15}.

Generally speaking, it is conceivable that the approaches based on optimal-control theory and shortcuts to adiabaticity can 
even be combined into a unified framework. This boils down to the fundamental open question as to whether it is possible to 
connect the Lewis-Riesenfeld invariants~\cite{Chen+:11} -- used for shortcuts to adiabaticity -- with the Pontryagin maximum 
principle~\cite{Pontryagin+:64} that forms the basis of optimal-control theory. If such a connection proves to be viable, this 
would allow one to combine the advantages of both approaches.

\section{Outlook: open-system effects} \label{outlook}
As hinted in Sec.~\ref{System_a} an all-encompassing approach to the gate-optimization 
problem at hand necessitates the inclusion of open-system effects, i.e., the unavoidable 
decoherence-induced noise. Here we provide a general assessment of this problem and briefly 
explain one possible approach for its quantitative treatment.

Regardless of the specific character of the qubit array and its environment (Markovian 
or non-Markovian), optimal-control-based gate synthesis with the inclusion of open-system 
effects is, generally speaking, computationally very expensive. This stems from the need to simulate 
quantum dynamics in a high-dimensional Hilbert space~\cite{Grace+:07,Floether+:12}. For instance, in 
Ref.~\cite{Floether+:12} such a study was carried out for small qubit systems with Heisenberg-type 
exchange coupling, which interact with either Markovian or non-Markovian environments. This study 
concluded that control fields optimized in the absence of the environment (closed system) remain the 
optimal ones in the Markovian case provided that the decoherence is sufficiently uniform and weak 
to be viewed as a perturbation of the unitary evolution. On the other hand, such pre-optimized 
fields were found to perform poorly in the non-Markovian case, thus underscoring the importance of 
an accurate characterization of the system-environment coupling for high-fidelity gate realizations.

While a full-fledged gate optimization in the open-system scenario is a rather difficult problem, 
a somewhat simpler task is to quantify how a gate-specific pulse sequence optimized for a closed system 
performs in the presence of decoherence-induced noise. This naturally entails the notion of the average 
state fidelity, which for a generic $N$-qubit system is defined as 
\begin{equation}
\bar{F}=2^{-N}\:\sum_{k}\sqrt{|\langle \psi_k|
\:\rho^{\textrm{fin}}_k\:|\psi_k\rangle|} \:.
\end{equation}
Here $|\psi_k\rangle$ ($k=1,\ldots,2^N$) are the (normalized) computational basis states of the 
system, while $\rho^{\textrm{fin}}_k$ is the density matrix at the end of a nonunitary evolution 
(i.e., at $t=t_g$, where $t_g$ is the time required for a high-fidelity realization of 
the concrete gate) that starts with the system in the pure state $|\psi_k\rangle$. In other words, 
$\rho^{\textrm{fin}}_k\equiv\rho(t=t_g)$, where $\rho(t)$ is the density matrix of the system which 
satisfies the initial condition $\rho(t=0)=|\psi_k\rangle\langle\psi_k|$. In the framework of 
the quantum operation formalism~\cite{NielsenChuangBook}, this density matrix can be written 
in the form of a sum over (time-dependent) Kraus matrices of the system~\cite{Kraus:71}.

The Kraus matrices of a qubit array are given by the tensor products of those representing individual qubits. 
To construct these single-qubit matrices one ought to adopt a specific model for the decoherence-induced 
noise. In one of the widely used models~\cite{Liu+:04}, a qubit is represented by the lowest two number states
of a linear harmonic oscillator and the environment as a collection of multimode oscillators. A qubit 
is subject to two noise processes, namely the amplitude and phase damping, each characterized by its 
own damping rate -- the respective inverses of the amplitude-relaxation- ($T_1$) and dephasing ($T_2$) 
times. The latter, usually similar in magnitude, are often assumed to be approximately the same and 
represented by the unique coherence time $T$. 

On quite general grounds, assuming that the decoherence-induced errors are mutually independent, 
the average state fidelity can be expected to be approximately given by $\bar{F}\approx F\:\exp(-t_g/T)$, 
where $F$ is the intrinsic fidelity. In cases where the achievable gate times are much shorter than the 
coherence time ($t_g\ll T$), the last expression simplifies to $\bar{F}\approx F\: [1-(t_g/T)]$. Unsurprisingly, 
such linear dependence of $\bar{F}$ on $t_g/T$ was predicted, for example, in a theoretical proposal for an 
avoided-crossing-based Toffoli and Fredkin gates in a system of three coupled SC transmon qubits~\cite{Zahedinejad++}.

As far as the system at hand is concerned, the characteristic times that we obtained for high-fidelity 
realizations of Toffoli and Fredkin gates are at most around $30\:J^{-1}$. For typical magnitudes of 
exchange-coupling constants in state-of-the-art SC- and spin-qubit systems (cf. Sec.~\ref{PiecewiseFiltered})
this amounts to the approximate gate times $t_g\sim 90 - 240$\:ns. On the other hand, typical coherence times in both of these 
classes of solid-state QC platforms are nowadays of the order of several tens-of-microseconds. Therefore, the 
condition $t_g\ll T$ is fulfilled in physical systems of relevance for the present investigation. In accordance 
with the reasoning mentioned above, this last conclusion also implies that one can expect to extract the linear 
dependence of $\bar{F}$ on $t_g/T$ in future studies that will take into account the open-system effects.

\section{Summary and Conclusions} \label{sumconclude}
To summarize, we investigated the feasibility of single-shot realizations of the Toffoli 
and Fredkin gates in qubit arrays with Heisenberg-type coupling between adjacent qubits. In 
doing so, we fully exploited the local controllability of this system, i.e., the 
fact that it is rendered completely controllable via a Zeeman-like control of a single 
actuator qubit. This control setting does not only reduce the burden of finding 
the optimal control fields -- by lowering their number -- but is also desirable 
because it alleviates the debilitating effects of decoherence. The present study incorporated 
two important practical issues of relevance for gate realizations: a finite-frequency range of 
realistic control fields and their leakage away from the actuator. It was demonstrated 
that none of these two ingredients presents an obstacle to realizing the Toffoli- and
Fredkin gates with high fidelities required for fault-tolerant quantum computing.

The synthesis of complex multi-qubit gates from single- and two-qubit building blocks
proved to be quite cumbersome. For example, four-qubit Toffoli gate employed in a recent 
implementation of Grover's search algorithm with trapped-ion qubits~\cite{Figgatt+:17} required as 
many as $11$ two-qubit gates and $22$ single-qubit gates. This fuels the need for alternative 
gate-synthesis approaches that avoid the use of such decompositions~\cite{Groenland+Schoutens:18}. 
The present work constitutes an attempt in this direction, specifically devoted to
systems with Heisenberg-type exchange interaction between adjacent qubits. In particular,
our findings regarding the efficient single-shot realization of the three-qubit 
Toffoli gate may facilitate future applications of this gate in measurement-free quantum 
error correction in this type of systems~\cite{TanamotoQECC,Sohn+:17} Likewise, the proposed
single-shot Fredkin gate may prove beneficial in the context of recently investigated
universal quantum computation utilizing continuous-variable bosonic cavity modes in three-dimensional 
circuit-QED architecture, where the central physical mechanism behind entangling such 
modes is an engineered exchange interaction~\cite{EngExchange}.

In conclusion, fast and accurate realizations of quantum gates remain one of the 
crucial ingredients towards attaining the overarching goal of universal quantum 
computation~\cite{PreskillNISQ:18}. The present work, which can be generalized 
to more complex (e.g., higher-dimensional) qubit networks~\cite{Arenz+Rabitz:18}, 
seems to indicate that the use of the single-shot approach could significantly alleviate 
the burden on control-generating hardware in future experimental realizations of multi-qubit 
gates. It will hopefully foster further experimental applications of this methodology. 

\begin{acknowledgments}
The author acknowledges useful discussions during previous collaborations 
on related topics with R. Heule, D. Burgarth, C. Bruder, and T. Tanamoto. 
\end{acknowledgments}


\begin{thebibliography}{57}
\expandafter\ifx\csname natexlab\endcsname\relax\def\natexlab#1{#1}\fi
\expandafter\ifx\csname bibnamefont\endcsname\relax
  \def\bibnamefont#1{#1}\fi
\expandafter\ifx\csname bibfnamefont\endcsname\relax
  \def\bibfnamefont#1{#1}\fi
\expandafter\ifx\csname citenamefont\endcsname\relax
  \def\citenamefont#1{#1}\fi
\expandafter\ifx\csname url\endcsname\relax
  \def\url#1{\texttt{#1}}\fi
\expandafter\ifx\csname urlprefix\endcsname\relax\def\urlprefix{URL }\fi
\providecommand{\bibinfo}[2]{#2}
\providecommand{\eprint}[2][]{\url{#2}}

\bibitem[{QCr()}]{QCreviews2013}
\bibinfo{note}{See, e.g., C. Monroe and J. Kim, Science ${\mathbf{339}}$, 1164
  (2013); M. H. Devoret and R. J. Schoelkopf, {\em ibid.} ${\mathbf{339}}$,
  1169 (2013); D. D. Awschalom, L. C. Bassett, A. S. Dzurak, E. L. Hu, and J.
  R. Petta, {\em ibid.} ${\mathbf{339}}$, 1174 (2013).}

\bibitem[{rec()}]{recentQCimport}
\bibinfo{note}{R. Barends {\em et al.}, Nature (London) {\bf 508}, 500 (2014); 
T. F. Watson {\em et al.}, {\em ibid.} {\bf 555}, 633 (2018).}

\bibitem[{Pre()}]{PreskillNISQ:18}
\bibinfo{note}{J. Preskill, Quantum {\bf 2}, 79 (2018).}

\bibitem[{Los()}]{Loss+DiVincenzo:98}
\bibinfo{note}{D. Loss and D. P. DiVincenzo, Phys. Rev. A {\bf 57}, 
 120 (1998).}

\bibitem[{Klo()}]{Kloeffel+Loss:13}
\bibinfo{note}{For a review, see C. Kloeffel and D. Loss, Annu.
  Rev. Condens. Matter Phys. {\bf 4}, 51 (2013); R. Hanson, L. P. Kouwenhoven, 
  J. R. Petta, S. Tarucha, and L. M. K. Vandersypen, Rev. Mod. Phys. 
  {\bf 79}, 1217 (2007).}

\bibitem[{Zaj()}]{Zajac+:16}
\bibinfo{note}{D. M. Zajac, T. M. Hazard, X. Mi, E. Nielsen, and J. R. Petta,
  Phys. Rev. Applied ${\mathbf{6}}$, 054013 (2016).}

\bibitem[{Vel()}]{Veldhorst+:15}
\bibinfo{note}{M. Veldhorst {\em et al.}, Nature (London) ${\mathbf{526}}$, 410
  (2015).}

\bibitem[{Ras()}]{Rassmussen+:18}
\bibinfo{note}{S. E. Rassmussen, K. S. Christensen, and N. T. Zinner, 
arXiv:1808.09881.}  

\bibitem[{DiV()}]{DiVincenzo+:00}
\bibinfo{note}{D. P. DiVincenzo, D. Bacon, J. Kempe, G. Burkard, and K. B.
  Whaley, Nature (London) {\bf 408}, 339 (2000).}

\bibitem[{Lev()}]{Levy:02}
\bibinfo{note}{J. Levy, Phys. Rev. Lett. {\bf 89}, 147902 (2002).}

\bibitem[{Enc()}]{EncodedUniversality}
\bibinfo{note}{D. Bacon, J. Kempe, D. A. Lidar, and K. B. Whaley, Phys. Rev.
  Lett. {\bf 85}, 1758 (2000).}

\bibitem[{Sch()}]{Schirmer++:08}
\bibinfo{note}{S. G. Schirmer, I. C. H. Pullen, and P. J. Pemberton-Ross, Phys.
  Rev. A ${\mathbf{78}}$, 062339 (2008).}

\bibitem[{\citenamefont{Wang et~al.}(2016)\citenamefont{Wang, Burgarth, and
  Schirmer}}]{Wang++:16}
\bibinfo{author}{\bibfnamefont{X.}~\bibnamefont{Wang}},
  \bibinfo{author}{\bibfnamefont{D.}~\bibnamefont{Burgarth}}, \bibnamefont{and}
  \bibinfo{author}{\bibfnamefont{S.~G.} \bibnamefont{Schirmer}},
  \bibinfo{journal}{Phys. Rev. A} \textbf{\bibinfo{volume}{94}},
  \bibinfo{pages}{052319} (\bibinfo{year}{2016}).

\bibitem[{Heu()}]{Heule+++}
\bibinfo{note}{R. Heule, C. Bruder, D. Burgarth, and V. M. Stojanovi\'c, Phys.
  Rev. A ${\mathbf{82}}$, 052333 (2010); Eur. Phys. J. D ${\mathbf{63}}$, 41
  (2011).}

\bibitem[{\citenamefont{Nielsen and Chuang}(2000)}]{NielsenChuangBook}
\bibinfo{author}{\bibfnamefont{M.~A.} \bibnamefont{Nielsen}} \bibnamefont{and}
  \bibinfo{author}{\bibfnamefont{I.~L.} \bibnamefont{Chuang}},
  \emph{\bibinfo{title}{{Q}uantum {C}omputation and {Q}uantum {I}nformation}}
  (\bibinfo{publisher}{Cambridge University Press},
  \bibinfo{address}{Cambridge}, \bibinfo{year}{2000}).

\bibitem[{Gre()}]{Green+:13}
\bibinfo{note}{T. J. Green, J. Sastrawan, H. Uys, and M. J. Biercuk, 
New J. Phys. {\bf 15}, 095004 (2013).}

\bibitem[{Con({\natexlab{a}})}]{ControlReviews}
\bibinfo{note}{For a comprehensive review, see S. J. Glaser 
{\em et al.}, Eur. Phys. J. D ${\mathbf{69}}$, 279 (2015).}

\bibitem[{Ash()}]{Ashhab+:12}
\bibinfo{note}{S. Ashhab, P. C. DeGroot, and F. Nori, Phys. Rev. A ${\mathbf{85}}$,
  052327 (2012).}  

\bibitem[{SCq()}]{SCqubitControl}
\bibinfo{note}{See, e.g., S. Montangero, T. Calarco, and R. Fazio, Phys. Rev.
  Lett. ${\mathbf{99}}$, 170501 (2007); P. Rebentrost and F. K. Wilhelm, Phys. Rev.
  B ${\mathbf{79}}$, 060507(R) (2009); M. Goerz, D. Reich, and C. Koch, 
  New J. Phys. ${\mathbf{16}}$, 055012 (2014); A. W. Cross and J. M. Gambetta, 
  Phys. Rev. A ${\mathbf{91}}$, 032325 (2015); E. Barnes, C. Arenz, A. Pitchford, 
  and S. E. Economou, Phys. Rev. B ${\mathbf{96}}$, 024504 (2017).}

\bibitem[{\citenamefont{Stojanovi{\'{c}}
  et~al.}(2012)\citenamefont{Stojanovi{\'{c}}, Fedorov, Wallraff, and
  Bruder}}]{StojanovicToffoli:12}
\bibinfo{author}{\bibfnamefont{V.~M.} \bibnamefont{Stojanovi{\'{c}}}},
  \bibinfo{author}{\bibfnamefont{A.}~\bibnamefont{Fedorov}},
  \bibinfo{author}{\bibfnamefont{A.}~\bibnamefont{Wallraff}}, \bibnamefont{and}
  \bibinfo{author}{\bibfnamefont{C.}~\bibnamefont{Bruder}},
  \bibinfo{journal}{Phys. Rev. B} \textbf{\bibinfo{volume}{85}},
  \bibinfo{pages}{054504} (\bibinfo{year}{2012}).

\bibitem[{Zah()}]{Zahedinejad++}
\bibinfo{note}{E. Zahedinejad, J. Ghosh, and B. C. Sanders, Phys. Rev. Lett.
  ${\mathbf{114}}$, 200502 (2015); Phys. Rev. Applied ${\mathbf{6}}$, 054005
  (2016).}

\bibitem[{Con({\natexlab{c}})}]{ControlNuclSpins}
\bibinfo{note}{C. D. Aiello and P. Cappellaro, Phys. Rev. A ${\mathbf{91}}$,
  042340 (2015); J. Zhang, D. Burgarth, R. Laflamme, and D. Suter, {\em ibid.}
  ${\mathbf{91}}$, 012330 (2015).}
  
\bibitem[{Tof()}]{ToffoliPapers:2017}
\bibinfo{note}{For recent examples, see J. K. Moqadam, G. S. Welter, and
  P. A. A. Esquef, Quantum Inf. Process. ${\mathbf{15}}$, 4501 (2016); A.
  Daskin and S. Kais, {\em ibid.} ${\mathbf{16}}$, 33 (2017); F. Holik, G.
  Sergioli, H. Freytes, R. Giuntini, and A. Plastino, {\em ibid.}
  ${\mathbf{16}}$, 1573 (2017); A. Devra, P. Prabhu, H. Singh, Arvind, and K.
  Dorai, {\em ibid.} ${\mathbf{17}}$, 67 (2018).}

\bibitem[{Mon()}]{Monz++:09}
\bibinfo{note}{T. Monz {\em et al.}, Phys. Rev. Lett. ${\mathbf{102}}$, 040501
  (2009).}

\bibitem[{Lan()}]{Lanyon++:09}
\bibinfo{note}{B. P. Lanyon {\em et al.}, Nat. Phys. ${\mathbf{5}}$, 134
  (2009).}

\bibitem[{Ree()}]{Reed++:12}
\bibinfo{note}{M. D. Reed, L. DiCarlo, S. E. Nigg, L. Sun, L. Frunzio, S. M.
  Girvin, and R. J. Schoelkopf, Nature (London) ${\mathbf{482}}$, 382 (2012).}
  
\bibitem[{Fred()}]{FredkinGate++}
\bibinfo{note}{R. B. Patel, J. Ho, F. Ferreyrol, T. C. Ralph, and G. J. Pryde,
Sci. Adv. ${\mathbf{2}}$, e1501531 (2016); T. Ono, R. Okamoto, M. Tanida, H. F. Hofmann,
and S. Takeuchi, Sci. Rep. ${\mathbf{7}}$, 45353 (2017).}

\bibitem[{Exch()}]{EngExchange}
\bibinfo{note}{Y. Y. Gao, B. J. Lester, K. Chou, L. Frunzio, M. H. Devoret, 
L. Jiang, S. M. Girvin, and R. J. Schoelkopf, arXiv:1806.07401.}

\bibitem[{NRc()}]{NRcBook}
\bibinfo{note}{W. H. Press, S. A. Teukolsky, W. T. Vetterling, and B. P. Flannery,
{\em Numerical Recipes in C: The Art of Scientific Computing} (Cambridge University 
Press, Cambridge, 1999).}

\bibitem[{Tor()}]{TornZilinskasBook}
\bibinfo{note}{A. T\"{o}rn and A. \v{Z}ilinskas, {\em Global Optimization},
  Lecture Notes in Computer Science, vol 350 (Springer, Berlin, 1989).}

\bibitem[{Hua()}]{ControlConceptual}
\bibinfo{note}{V. Jurdjevi\'c and H. J. Sussmann, J. Differ. Equations 
      ${\mathbf{12}}$, 313 (1972); G. M. Huang, T. J. Tarn, and J. W. Clark, 
      J. Math. Phys. ${\mathbf{24}}$, 2608 (1983); V. Ramakrishna and H. Rabitz, 
      Phys. Rev. A ${\mathbf{54}}$, 1715 (1996).}

\bibitem[{\citenamefont{D'Alessandro}(2008)}]{D'AlessandroBook}
\bibinfo{author}{\bibfnamefont{D.}~\bibnamefont{D'Alessandro}},
  \emph{\bibinfo{title}{Introduction to {Q}uantum {C}ontrol and {D}ynamics}}
  (\bibinfo{publisher}{Taylor \& Francis}, \bibinfo{address}{Boca Raton},
  \bibinfo{year}{2008}).

\bibitem[{\citenamefont{Pfeifer}(2003)}]{PfeiferBook}
\bibinfo{author}{\bibfnamefont{W.}~\bibnamefont{Pfeifer}},
  \emph{\bibinfo{title}{The {L}ie {A}lgebras $su({N})$: {A}n {I}ntroduction}}
  (\bibinfo{publisher}{Birkh{\"{a}}user}, \bibinfo{address}{Basel},
  \bibinfo{year}{2003}).

\bibitem[{Loc()}]{LocalPlastina}
\bibinfo{note}{See, e.g., S. Lorenzo, T. J. G. Apollaro, A. Sindona,
  and F. Plastina, Phys. Rev. A ${\mathbf{87}}$, 042313 (2013).}

\bibitem[{Pio()}]{PioroLadriere+:08}
\bibinfo{note}{M. Pioro-Ladriere, T. Obata, Y. Tokura, Y.-S. Shin, T. Kubo, K.
  Yoshida, T. Taniyama, and S. Tarucha, Nat. Phys. ${\mathbf{4}}$, 776 (2008).}
  
\bibitem[{Wen()}]{Wendin:17}
\bibinfo{note}{For an up-to-date review, see G. Wendin, Rep. Prog. Phys. 
{\bf 80}, 106001 (2017).}
  
\bibitem[{Faz()}]{Fazio+Zant:01}
\bibinfo{note}{R. Fazio and H. van der Zant, Phys. Rep.
  ${\mathbf{355}}$, 235 (2001).} 
  
\bibitem[{Rei()}]{Reilly:15}
\bibinfo{note}{See, e.g., D. J. Reilly, npj Quantum Inf. {\bf 1}, 15011 (2015).}  

\bibitem[{vdi()}]{vanDijk+:18}
\bibinfo{note}{J. P. G. van Dijk, E. Charbon, and F. Sebastiano, arXiv:1811.01693.}  

\bibitem[{Rya()}]{Ryan+:17}
\bibinfo{note}{C. A. Ryan, B. R. Johnson, D. Riste, B. Donovan, and T. A. Ohki, 
Sci. Rev. Inst. {\bf 88}, 104703 (2017).} 

\bibitem[{\citenamefont{Lloyd}(1995)}]{Lloyd:95}
\bibinfo{author}{\bibfnamefont{S.}~\bibnamefont{Lloyd}},
  \bibinfo{journal}{Phys. Rev. Lett.} \textbf{\bibinfo{volume}{75}},
  \bibinfo{pages}{346} (\bibinfo{year}{1995}).

\bibitem[{Wea()}]{Weaver:00}
\bibinfo{note}{N. Weaver, J. Math. Phys. ${\mathbf{41}}$, 240 (2000).}  

\bibitem[{Luc()}]{Lucarelli:18}
\bibinfo{note}{D. Lucarelli, Phys. Rev. A ${\mathbf{97}}$, 062346 (2018).}

\bibitem[{Ste()}]{Steane:03}
\bibinfo{note}{A. M. Steane, Phys. Rev. A ${\mathbf{68}}$, 042322 (2003).}

\bibitem[{Kni()}]{KnillThreshold}
\bibinfo{note}{E. Knill, Nature (London) ${\mathbf{434}}$, 39 (2005).}

\bibitem[{Hig()}]{HighThreshold}
\bibinfo{note}{R. Raussendorf and J. Harrington, Phys. Rev. Lett.
  ${\mathbf{98}}$, 190504 (2007); S. D. Barrett and T. M. Stace, {\em ibid.}
  ${\mathbf{105}}$, 200502 (2010); D. S. Wang, A. G. Austin, and L. C. L.
  Hollenberg, Phys. Rev. A ${\mathbf{83}}$, 020302(R) (2011).}

\bibitem[{She()}]{Shende+Markov:09}
\bibinfo{note}{V. V. Shende and I. L. Markov, Quantum Inf. Comput.
  ${\mathbf{9}}$, 0461 (2009).}
  
\bibitem[{Aar()}]{AarhusCircuitGates:18}
\bibinfo{note}{T. B\ae{a}kkegaard, L. B. Kristensen, N. J. S. Loft, C. K. Andersen, 
D. Petrosyan, and N. T. Zinner, arXiv:1802.04299.}

\bibitem[{Zan()}]{Zanardi+Rasetti:99}
\bibinfo{note}{P. Zanardi and M. Rasetti, Phys. Lett. A {\bf 264}, 94 (1999).}

\bibitem[{Sjo()}]{Sjoqvist+:12}
\bibinfo{note}{E. Sj\"{o}qvist, D. M. Tong, L. M. Andersson, B. Hessmo, 
M. Johansson, and K. Singh, New J. Phys. {\bf 14}, 103035 (2012).}

\bibitem[{Fen()}]{Feng+:13}
\bibinfo{note}{G. Feng, G. F. Xu, and G. L. Long, Phys. Rev. Lett. {\bf 110}, 
190501 (2013).}

\bibitem[{Che()}]{Chen+:18}
\bibinfo{note}{T. Chen, J. Zhang, and Z.-Y. Xue, Phys. Rev. A {\bf 98}, 
 052314 (2018).}

\bibitem[{Che()}]{Chen+:11}
\bibinfo{note}{X. Chen, E. Torrontegui, and J. G. Muga, 
Phys. Rev. A {\bf 83}, 062116 (2011).}

\bibitem[{Iba()}]{Ibanez+:12}
\bibinfo{note}{S. Ib\'{a}\~{n}ez, X. Chen, E. Torrontegui, J. G. Muga, and A. Ruschhaupt, 
Phys. Rev. Lett. {\bf 109}, 100403 (2012).}

\bibitem[{Zha()}]{Zhang+:15}
\bibinfo{note}{J. Zhang, T. H. Kyaw, D. M. Tong, E. Sj\"{o}qvist, and L.-C. Kwek, 
Sci. Rep. {\bf 5}, 18414 (2015).}

\bibitem[{Pon()}]{Pontryagin+:64}
\bibinfo{note}{L. S. Pontryagin, V. G. Bol'tanskii, R. S. Gamkrelidze, and E. F. Mischenko, 
{\em The Mathematical Theory of Optimal Processes} (Pergamon Press, New York, 1964).}

\bibitem[{Gra()}]{Grace+:07}
\bibinfo{note}{M. Grace, C. Brif, H. Rabitz, I. A. Walmsley, 
R. L. Kosut, and D. Lidar, J. Phys. B: At. Mol. Opt. Phys. 
${\mathbf{40}}$, S103 (2007).}

\bibitem[{Flo()}]{Floether+:12}
\bibinfo{note}{F. F. Floether, P. de Fouquieres, and S. G. Schirmer, 
New J. Phys. ${\mathbf{14}}$, 073023 (2012).}

\bibitem[{Kra()}]{Kraus:71}
\bibinfo{note}{K. Kraus, Ann. Phys. (N.Y.) ${\mathbf{64}}$, 311 (1971).}

\bibitem[{Liu()}]{Liu+:04}
\bibinfo{note}{Y.-x. Liu, S. K. \"{O}zdemir, A. Miranowicz, and N. Imoto, 
  Phys. Rev. A {\bf 70}, 042308 (2004).}

\bibitem[{Fig()}]{Figgatt+:17}
\bibinfo{note}{C. Figgatt, D. Maslov, K. A. Landsman, N. M. Linke, S. Debnath,
  and C. Monroe, Nat. Commun. ${\mathbf{8}}$, 1918 (2017).}

\bibitem[{\citenamefont{Groenland and
  Schoutens}(2018)}]{Groenland+Schoutens:18}
\bibinfo{author}{\bibfnamefont{K.}~\bibnamefont{Groenland}} \bibnamefont{and}
  \bibinfo{author}{\bibfnamefont{K.}~\bibnamefont{Schoutens}},
  \bibinfo{journal}{Phys. Rev. A} \textbf{\bibinfo{volume}{97}},
  \bibinfo{pages}{042321} (\bibinfo{year}{2018}).
  
\bibitem[{Are()}]{Arenz+Rabitz:18}
\bibinfo{note}{C. Arenz and H. Rabitz, Phys. Rev. Lett. ${\mathbf{120}}$, 220503 (2018).}  

\bibitem[{Tan()}]{TanamotoQECC}
\bibinfo{note}{See, e.g., T. Tanamoto, V. M. Stojanovi\'c, C. Bruder, and D.
  Becker, Phys. Rev. A ${\mathbf{87}}$, 052305 (2013); T. Tanamoto, {\em ibid.}
  ${\mathbf{88}}$, 062334 (2013).}

\bibitem[{Soh()}]{Sohn+:17}
\bibinfo{note}{I. Sohn, S. Tarucha, and B.-S. Choi, Phys. Rev. A
  ${\mathbf{95}}$, 012306 (2017).}
  
\end{thebibliography}
\end{document}